\documentclass[a4paper,11pt]{article}
\pdfoutput=1
cm \voffset = -2truecm

\usepackage{graphicx}
\usepackage{amsmath}
\usepackage{amssymb}
\usepackage{latexsym}
\usepackage[colorlinks]{hyperref}
\usepackage{color}
\usepackage{cite}
\usepackage{makeidx}
\usepackage{float}
\usepackage{multirow}

\restylefloat{figure}
\newcommand{\bra}{\begin{array}}
\newcommand{\era}{\end{array}}
\newcommand{\beq}{\begin{equation}}
\newcommand{\eeq}{\end{equation}}
\newcommand{\bqr}{\begin{eqnarray}}
\newcommand{\eqr}{\end{eqnarray}}

\def\BC{\bb C}
\def\_\BC{\bbi C}

\def\( {\left(}
   \def\) {\right)}
\def\[ {\left[}
\def\] {\right]}
\def\no2 {{\textstyle{n\over 2}}}

\newcommand{\al}{\alpha}

\newcommand{\lb}{\label}

\begin{document}
\begin{titlepage}
\setcounter{page}{1}
\renewcommand{\thefootnote}{\fnsymbol{footnote}}

\begin{flushright}
\end{flushright}

\vspace{5mm}
\begin{center}

{\Large \bf {Transmissions in Graphene
through Double Barriers \\
and Periodic Potential}}

\vspace{5mm}
{\bf Miloud Mekkaoui}$^{a}$, {\bf El Bou\^{a}zzaoui Choubabi}$^{a}$,
 {\bf Ahmed Jellal\footnote{\sf ajellal@ictp.it --
a.jellal@ucd.ac.ma}}$^{a,b}$ and {\bf Hocine Bahlouli}$^{b,c}$

\vspace{5mm}

{$^{a}$\em Theoretical Physics Group,  
Faculty of Sciences, Choua\"ib Doukkali University},\\
{\em PO Box 20, 24000 El Jadida, Morocco}

{$^b$\em Saudi Center for Theoretical Physics, Dhahran, Saudi Arabia}

{$^c$\em Physics Department,  King Fahd University
of Petroleum $\&$ Minerals,\\
Dhahran 31261, Saudi Arabia}



\vspace{3cm}

\begin{abstract}

Transmission of Dirac fermions through a chip of graphene
under the effect of magnetic field and a time vibrating double
barrier with frequency $w$ is investigated.
Quantum interference within the oscillating barrier has an
important effect on quasi-particles tunneling. A combination
of both a time dependent potential and a magnetic field
generate physical states whose energy is double quantified by
the pair of integers $(n, l)$ with high degeneracy. The large number
of modes that exist in the energy spectrum presents
a colossal difficulty in numerical computations. Thus we were obliged
to make a truncation and limit ourselves to the central $( n = 0 )$ and
two adjacent side band ($n=\pm 1$).

\vspace{3cm}



\end{abstract}
\end{center}
\end{titlepage}


\section{ Introduction}

{
Graphene \cite{Geim} is a single layer of carbon atoms arranged into a planar honeycomb
lattice. Since its experimental realization in 2004 \cite{Novoselov1} this system has attracted
a considerable attention from both experimental and theoretical researchers . This is because of
its unique and outstanding mechanical, electronic, optical, thermal and chemical properties \cite{Castro}.
Most of these marvelous properties are due
to the apparently relativistic-like nature of its carriers, electrons behave as massless Dirac fermions in graphene systems.
In fact starting from the original tight-binding Hamiltonian describing  graphene it has been shown theoretically
that the low-energy excitations of graphene appear to be massless chiral Dirac fermions. Thus, in the continuum
limit one can analyze the crystal properties using the formalism of quantum electrodynamics in (2+1)-dimensions.
This similarity between condensed matter physics and quantum electrodynamics (QED) provides the opportunity
to probe many physical aspects proper to high energy physics phenomena in condensed matter systems. Thus, in this regard,
graphene can be considered as a test-bed laboratory for high energy relativistic quantum phenomena.

Quantum transport in periodically driven quantum systems is an important subject not only of academic value but also
for device and optical applications. In particular quantum interference within an oscillating time-periodic electromagnetic field
gives rise to additional sidebands at energies $\epsilon + l\hbar \omega$ $ (l=0,\pm1,\cdots)$ in the transmission probability originating from
the fact that electrons exchange energy quanta $\hbar \omega $ carried by photons of the oscillating field, $\omega $ being the frequency
of the oscillating field. The standard model in this context is that of a time-modulated scalar potential in a finite region of space.
It was studied earlier by Dayem and Martin \cite{Dayem} who provided the experimental evidence of photon assisted tunneling in experiments on superconducting
films under microwave fields. Later on  Tien and Gordon \cite{Tien} provided the first theoretical explanation of these experimental
observations. Further theoretical studies were performed later by many research groups, in particular Buttiker investigated the
barrier traversal time of particles interacting with a time-oscillating barrier \cite{Buttiker}.
{Wagner  \cite{Wagner-1} gave a detailed
treatment on photon-assisted tunneling through a strongly driven double
barrier tunneling diode and studied the transmission probability of
electrons traversing a quantum well subject to a harmonic driving force
\cite{Wagner-2} where transmission side-bands have been predicted.
Grossmann \cite{Grossmann}, on the other hand, investigated the
tunneling through a double-well perturbed by a monochromatic driving
force which gave rise to unexpected modifications in the tunneling
phenomenon.}

In \cite{ahsan} the authors studied  the chiral tunneling through a harmonically driven potential barrier in a graphene monolayer.
Because the charge carriers in their system are massless they described the tunneling effect as the Klein tunneling with high anisotropy.
For this, they determined the transmission probabilities for the central band and sidebands in terms of the incident angle of
the electron beam. Subsequently, they investigated the transmission probabilities for varying width, amplitude and frequency of the
oscillating barrier. They conclude that the perfect transmission for normal incidence, which has been reported for a static barrier,
persists for the oscillating barrier which is a manifestation of Klein tunneling in a time-harmonic potential.

The growing experimental interest in studying optical properties of electron transport in graphene subject to strong laser fields \cite{Jiang1}
motivated the recent upsurge in theoretical study of the effect of time dependent periodic electromagnetic field on electron spectra.
Recently it was shown that laser fields can affect the electron density of states and consequently the electron transport properties \cite{Calvo}.
{Electron transport in graphene generated by laser
irradiation was shown to result in subharmonic resonant enhancement
\cite{San-Jose}.
The analogy between spectra of Dirac fermions in laser fields and the
energy spectrum in graphene superlattice formed by static one
dimensional periodic potential was recently performed \cite{Savelev-1}. In
graphene systems resonant enhancement of both electron backscattering
and currents across a scalar potential barrier of arbitrary space and
time dependence was investigated in \cite{Savelev-2} and resonant
sidebands in the transmission due to a time modulated potential
was studied recently in graphene \cite{Liu}. The fact that an applied
oscillating field can
result in an effective mass or equivalently a dynamic gap was confirmed
in recent studies \cite{Fistul}. Adiabatic quantum pumping of a graphene
devise with two oscillating electric barriers was considered
\cite{Evgeny}. A Josephson-like current was predicted for several time
dependent scalar potential barriers placed upon a monolayer of graphene
\cite{Savelev-3}. Stochastic resonance like phenomenon
\cite{Gammaitoni} was predicted for transport phenomena in disordered
graphene nanojunctions \cite{Jiang}. Further study showed that
noise-controlled effects can be induced due to the interplay between
stochastic and relativistic dynamics of charge carriers in graphene
\cite{Pototsky}.}

Very recently, we have analyzed the energy spectrum of graphene sheet with a
single barrier structure having a time periodic oscillating height in the presence of a magnetic field
\cite{Mekkaoui2}. The corresponding transmission was studied as a function of the
energy and the potential parameters. We have shown that quantum interference within
the oscillating barrier has an important effect on quasiparticle tunneling.
In particular the time-periodic electromagnetic field generates additional sidebands
at energies $\epsilon + l\hbar \omega (l=0,\pm 1, \cdots)$ in the transmission
probability originating from the photon absorption or emission within the oscillating barrier.
Due to numerical difficulties in truncating the resulting coupled channel equations
we have limited ourselves to low quantum channels, i.e. $l=0,\pm 1.$

We extend our previous work \cite{Mekkaoui2} to  consider monolayer graphene sheet
in the presence of magnetic field but with double barriers
 along
the $x$-direction while the carriers are free in the $y$-direction.
The barrier height oscillates sinusoidally around an average value
$V_{j}$ with oscillation amplitude $U_{j}$ and frequency $\omega$.
The spectral solutions are obtained in the five  regions forming our sheet
as functions of different physical parameters. These are used to calculate
the current density and therefore evaluate
the transmission probability for the central band and
close by sidebands as a function of the potential parameters and
incident angle of the particles. We present our numerical results and discuss their
implications for low quantum channels.

The manuscript is organized as follows. In section 2, we present our theoretical model by defining the governing Hamiltonian
 and setting the applied potentials and external magnetic field. We solve the resulting eigenvalue equations
to obtain the  solutions of the energy spectrum for the five regions composing our system in section 3.
Using the boundary conditions as well as the current density we exactly  determine the
transmission probability in section 4. Our main results and comparisons with existing literature
will be presented in section 5. We conclude by summarizing our main results in the last section.

\section{Hamiltonian of the system}

Consider a two-dimensional system of Dirac fermions forming a
sheet graphene. This sheet is subject to a vibrating double barrier potential
in addition to a mass term and an externally applied magnetic field as shown in Figure \ref{fig11}.
Particles and antiparticles moving respectively in the positive and negative energy regions with the
tangential component of the wave vector along the
$x$-direction have translation invariance in the
$y$-direction. Dirac  fermions move through a monolayer
graphene and scatter off a double barrier potential whose
height is oscillating sinusoidally around $V_{j}$ with
amplitude $U_{j}$ and frequency $\omega$. The carriers are also
subject to a magnetic field perpendicular to the graphene sheet
$\textbf{B}= B(x, y)\textbf{e}_{z}$ and a mass term is added to a
vector potential coupling.  Dirac  fermions with energy $E$ are
incident with an angle $\phi_{1}^{1}$ with respect to the $x$-axis,
the conservation of energy allows the appearance of an infinite number of modes with levels
$E+ m\hbar \omega$ $(m=0, \pm 1, \pm 2 \cdots)$. The Hamiltonian
governing  the system is composed by two independent terms
$(H_{0},H_{1}) $
\begin{equation}
H=H_{0}+H_{1} \lb{seq0}
\end{equation}
where the first part is
\begin{equation} \lb{ch1}
H_{0}=v_{F} \sigma \cdot \left(-i\hbar\nabla+
\frac{e}{c}\textbf{A}(x,y)\right)+V(x){\mathbb
I}_{2}+\kappa\Theta\left(d_{1}^{2}-x^{2}\right)\sigma_{z}
\end{equation}
and the oscillating barrier potential is defined in each scattering region by (see Figure \ref{fig11})
\begin{equation}
H^j_{1}=U_{j}\cos\left(\omega t+\delta_{j}\Theta(x)\right).
\end{equation}
The Hamiltonian $H_{1}$ describes the harmonic time dependence of
the barrier height, $\kappa$ is the mass term, $\upsilon_{F}$ the
Fermi velocity, $ \sigma =(\sigma_{x}, \sigma_{y})$ are the usual
Pauli matrices, phase difference $\delta_{j}$,  ${\mathbb I}_{2}$
the $2 \times 2$ unit matrix, the electrostatic potential $V(x)=V_{j}$ in each scattering region and
the magnetic field $B(x, y)= B(x)$.
Adopting  the Landau gauge which allows the vector potential to be
of the form $A = (0,A_{y}(x))$ with $\partial_{x}A_{y}(x)=
B(x)$, the transverse momentum $p_{y}= - i\partial_{y}$ is thus
conserved. The magnetic field $ \textbf{B} = B_{0}\textbf{e}_{z}$
(with constant $B_{0}$) within the strip $ |x|\leq d_{1}$ but $B=0$
elsewhere
\begin{equation}
B(x,y)= B_{0}\Theta(d_{1}^{2}-x^{2})
\end{equation}
with the Heaviside step function $\Theta$
\begin{equation}
\Theta(x)=\left\{%
\begin{array}{ll}
    1, & \hbox{$x>0$} \\
    0, & \hbox{otherwise.} \\
\end{array}%
\right.
\end{equation}
The  static square potential barrier $V(x)$ is defined by its constant value $V_{j}$
in each region, similarly for the amplitude of the
oscillating potential $U_{j}$
\begin{equation}
V(x)=V_{j}=
\left\{%
\begin{array}{ll}
    V_{2}, & \hbox{$d_{1}\leq |x|\leq d_{2}$} \\
    V_{3}, & \hbox{$ |x|\leq d_{1}$} \\
    0, & \hbox{otherwise} \\
\end{array}%
\right.,\qquad U_{j}=
\left\{%
\begin{array}{ll}
    U_{2}, & \hbox{$-d_{2}\leq x\leq -d_{1}$} \\
    U_{4}, & \hbox{$d_{1}\leq x\leq d_{2}$} \\
    0, & \hbox{otherwise} \\
\end{array}%
\right.
\end{equation}
where the index $j = 1, 2, 3, 4, 5$ denotes each scattering region as shown
in Figure \ref{fig11}.

\begin{figure}[h]
\centering
\includegraphics[width=10
cm,height=5cm]{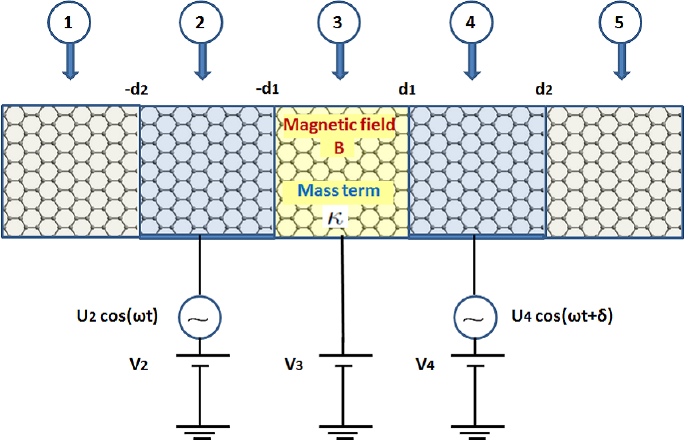}\\
 \caption{\sf{ (Color online)
   Schematic of  a graphene monolayer in the presence of an oscillating potential
   and a magnetic field. Different scattering regions are indicated by an integer j=1,2,3,4,5.}}\lb{fig11}
\end{figure}

Concerning the applied magnetic field, it is a constant and uniform
magnetic field ${\bf B}$ perpendicular to the graphene sheet but confined to a strip of width $2d$.
Due to incommensurate effect and interaction with substrate, graphene can develop a mass term in the Hamiltonian.
The vector potential that generates our magnetic field can be chosen of the following form
\begin{equation}
\qquad A_{y}(x)=\frac{c}{el_{B}^{2}}\left\{%
\begin{array}{ll}
    -d_{1}, & \hbox{$x<d_{1}$} \\
    x, & \hbox{$|x|< d_{1}$} \\
    d_{1}, & \hbox{$x>d_{1}$} \\
\end{array}%
\right.
\end{equation}
with the magnetic length defined by $l_{B}
=\sqrt{\frac{1}{B_{0}}}$  in the unit system $(\hbar= c = e =
1)$.

\section{Spectral solutions}

We emphasize that the system Hamiltonian \eqref{seq0} is composed of two sub-Hamiltonian,
$H_1$ plays the role of a perturbation term with respect to $H_0$.
The independence of these Hamiltonians leads to their
commutativity $[H_0,H_1]=0$ and therefore the corresponding
eigenspinors $\psi$ are the tensor product
of two eigenspinor $\psi_0$ and $\psi_1$  associated with $H_0$ and
$H_1$, respectively i.e.
$\psi(x , y, t) 
=\psi(x, y)\psi(t)$
and the eigenvalue of $H$ is the sum of eigenvalues
$E = E_0 + E_1.$
The eigenspinor $\psi$ of the system obeys the 
equation
\begin{equation}
H\psi(x,y,t)=  i \partial_t \psi_j(x,y,t)
\end{equation}
which can be written as
\begin{equation}
\left[E_0+U_{j}\cos\left(\omega
t+\delta_{j}\Theta(x)\right)\right]\psi_j(x,y,t)= i \partial_t \psi_j(x,y,t).
\end{equation}
The integration between $t_0 = 0$ and $t$ gives
\begin{eqnarray}
 \psi_j(x,y,t) 
= \psi_j(x,y,0)e ^{-i E_0 t} e ^{-i U_{j}\sin\left(\omega
t+\delta_{j}\Theta(x)\right) / \omega}
\end{eqnarray}
where
the last term  is in the form of $e^{i
\alpha_{j} \sin \Phi}$, which can be expanded into trigonometric series
as
\begin{equation}
  e^{i \alpha_{j}
\sin \Phi}=\sum_{m=-\infty}^{+\infty}J_m (\alpha_{j})e ^{i m\Phi}
\end{equation}
with $\alpha_{j}=\frac{U_j}{\omega}$ and $\Phi= \omega
t+\delta_{j}\Theta(x)$. Hence finally we obtain
\begin{eqnarray}\lb{ch0}
 \psi_j(x,y,t)= \psi_j(x,y,0)e ^{-i E_0 t}\sum_{m=-\infty}^{+\infty}J_m (\alpha_{j})e ^{i m(\omega t+ \delta)}
\end{eqnarray}
and $C_{m}=J_{m}\left(\alpha_{j}\right)$ satisfies the  recurrence relation
\begin{equation}
2mC_{m}=\alpha_{j}\left(C_{m+1}+C_{m-1}\right)
\end{equation}
where $J_{m}$ is the $m$-th order Bessel function of the first kind.
Using these eigenspinors we readily determine the total energy
from \eqref{ch0} to be
\begin{equation}
E=E_0+m\omega.
\end{equation}
Taking into account energy conservation, the wave packet that describes our carrier in
the $j$-th region can be expressed as a linear combination
of wave functions at energies $E_{0}+l\omega$ $(l=0,\pm 1,\cdots)$. This is
\begin{equation}\lb{ch2}
\psi_{ j}(x,y,t)=e^{ik_{y}y}\sum^{m,l=+\infty}_{m,l=-\infty}
\psi_{j}^{l}(x,y)
J_{m-l}\left(\alpha_{j}\right)e^{-i(m-l)\delta_{j}}e^{-iv_{F}(\epsilon+m\varpi)t}
\end{equation}
where we have set $\epsilon=\frac{E_0}{ \upsilon_{F}}$ and
$\varpi=\frac{\omega}{ \upsilon_{F}}$. Subsequently, the spinor $\psi_{\sf j}^{l}(x,y)$
will be determined in each region $j$.

The Dirac eigenvalue  equation in the absence of oscillating
potential for the spinor $\psi(x, y)=(\psi_{+},\psi_{-})^{T}$ at
energy $E_0$ reads
\begin{equation}
H_{0}\psi(x,y)= E_0 \psi(x,y).
\end{equation}
Using the explicit form of $H_0$ given by \eqref{ch1} we find
\begin{equation}\lb{ih00}
\left(%
\begin{array}{cc}
  v_{j}+\mu & p_{jx}-ip_{y}-i A(x) \\
  p_{jx}+ip_{y}+i A(x) & v_{j}-\mu \\
\end{array}%
\right)\left(%
\begin{array}{c}
  \psi_{+} \\
  \psi_{-} \\
\end{array}%
\right)=\epsilon\left(%
\begin{array}{c}
  \psi_{+} \\
  \psi_{-} \\
\end{array}%
\right)
\end{equation}
where $v_j=\frac{V_j}{v_F}$ and $\mu=\frac{\kappa}{v_F}$ .
Due to the translational invariance along  the
$y$-direction, the two-component pseudospinor can be written as
$\psi_{\pm}(x, y)= \varphi_{\pm}(x)e^{ik_{y}y}$.
In region $j$ = 1, 2, 4 and 5,
we easily obtain the following two linear differential equations
\begin{eqnarray}
  && \left (p_{jx}-p_{y}-i A(x) \right)\varphi_{-}=(\epsilon-v_{j}) \varphi_{+}\\
&&   \left (p_{jx}+p_{y}+i A(x)
\right)\varphi_{+}=(\epsilon-v_{j}) \varphi_{-}.
\end{eqnarray}
In accordance with \eqref{ch2}, the general solution in the $j$-th scattering region reads as
\begin{eqnarray}
\psi_{\sf
j}(x,y,t) &=&e^{ik_{y}y}\sum^{m,l=+\infty}_{m,l=-\infty}\left[a_{l}^{j}\left(
\begin{array}{c}
1 \\
 z_{l}^{j}\end{array}\right)e^{ik^{j}_{l} (x-x_{j})
 }+b_{l}^{j}\left(
\begin{array}{c}
1 \\
 -\frac{1}{z_{l}^{j}}\end{array}\right)e^{-ik^{j}_{l} (x-x_{j})}\right]\nonumber
\\
&&  \times J_{m-l}\left(\alpha_{j}\right)e^{-i(m-l)\delta_{j}}e^{-iv_{F}(\epsilon+m\varpi)t}
\end{eqnarray}
where
$s_{l}^{j}=\mbox{sgn}(\epsilon+l \varpi-v_{j})$, the sign again
refers to conduction and valence bands, $x_{j}$ are the
positions of the interfaces (Figure \ref{fig11}): $x_{1}=x_{2}=-d_{2}$,
$x_{3}=-d_{1}$, $x_{4}=d_{1}$, $x_{5}=d_{2}$. Note that,  outside the barrier regions where the
modulation amplitude is $\alpha_{j} = 0$ we have the function $J_{m-l}
\left(\alpha_{j}\right)=\delta_{m,l}$. The wave vector is given by
\begin{equation}
k_{l}^{j}=s_{l}^{j}\sqrt{\left(\epsilon-v_{j}+l\varpi\right)^{2}-\left(k_{y}+\frac{d}{l^{2}_{B}}\right)^{2}}
\end{equation}
which leads to the corresponding eigenvalues
\begin{equation}
\epsilon-v_{j}+l\varpi=s_{l}^{j}
\sqrt{\left(k_{l}^{j}\right)^{2}+\left(k_{y}+\frac{d}{l^{2}_{B}}\right)^{2}}
\end{equation}
with the magnetic length defined by $l_{B}=\sqrt{1/B_{0}}$ and the complex parameter $z_{l}^{j}$ is
\begin{equation}
z_{l}^{j}=s_{l}^{j}\frac{k^{j}_{l}
+i\left(k_{y}+\frac{d}{l^{2}_{B}}\right)}{\sqrt{\left(k_{l}^{j}\right)^{2}
+\left(k_{y}+\frac{d}{l^{2}_{B}}\right)^{2}}}=s_{l}^{j}
e^{\textbf{\emph{i}}\phi_{l}^{j}}
\end{equation}
 $\phi_{l}^{j}=\tan^{-1}(k_{y}/k_{l}^{j})$ and the parameter
$d$ is defined by
\begin{equation}
d=\left\{%
\begin{array}{ll}
    d_{1}, & \hbox{$x<-d_{1}$} \\
    -d_{1}, & \hbox{$x>d_{1}$.} \\
\end{array}%
\right.
\end{equation}

Let us proceed to write down the solution in the intermediate
zone ${j= 3}$ $(-d_{1}<x<d_{1})$ containing the mass term in addition to a
perpendicular magnetic field. To diagonalize the
corresponding Hamiltonian we introduce the usual boson
operators
\begin{eqnarray}
a_{l}=\frac{l_{B}}{\sqrt{2}} \left(\partial_{x}+k_{y}+\frac{x}{l_{B}^{2}} \right), \qquad
a_{l}^{\dagger}=\frac{l_{B}}{\sqrt{2}}
\left(-\partial_{x}+k_{y}+\frac{x}{l_{B}^{2}} \right)
\end{eqnarray}
which satisfy the commutation relation $\left[a_{l},
a_{k}^{\dagger}\right]=\delta_{lk}$.
In terms of $a_{l}$ and $a_{l}^{\dagger}$, equation \eqref{ih00}
reads
\begin{equation}
 \left(%
\begin{array}{cc}
  v_{3}+\mu& -i\frac{\sqrt{2}}{l_{B}}a_{l} \\
  i\frac{\sqrt{2}}{l_{B}}a_{l}^{\dagger}  &  v_{3}-\mu \\
\end{array}%
\right)\left(%
\begin{array}{c}
  \varphi_{l,1} \\
  \varphi_{l,2} \\
\end{array}%
\right)=(\epsilon+l\varpi)\left(%
\begin{array}{c}
  \varphi_{l,1} \\
  \varphi_{l,2}\\
\end{array}%
\right)
\end{equation}
or in its explicit form
\begin{eqnarray}
 && -i\frac{\sqrt{2}}{l_{B}} a_{l}\varphi_{l,2}=\left(\epsilon+l\varpi-v_{3}-\mu \right)\varphi_{l,1} \lb{feq}\\
  && i\frac{\sqrt{2}}{l_{B}}a_{l}^{\dagger}\varphi_{l,1}=\left(\epsilon+l\varpi-v_{3}+\mu \right)\varphi_{l,2} \lb{seq}.
\end{eqnarray}
Combining the above equations, we obtain 
for $\varphi_{l,1}$
\begin{equation}
\left(\left(\epsilon+l\varpi-v_{3}\right)^{2}-\mu^{2}\right)\varphi_{l,1}=\frac{2}{l_{B}^{2}}
a_{l} a_{l}^{\dagger}\varphi_{l,1}.
\end{equation}
It is clear that $\varphi_{l,1}$ is an eigenstate of the number
operator $\widehat{N}=a_{l}^{\dagger}a_{l}$ and therefore we
identify $\varphi_{l,1}$ with the eigenstates of the harmonic
oscillator $|n-1\rangle$, namely
\begin{equation}
 \varphi_{l,1} \sim \mid n-1\rangle
\end{equation}
and the associated eigenvalues are 
\begin{equation}
\epsilon-v_{3}+l\varpi=\pm\frac{1}{l_{B}}\sqrt{\left(\mu
l_{B}\right)^{2}+2n}.
\end{equation}
Finally, the solution in region $j=3$  can be expressed in accordance with equation \eqref{ch2}, as follows 
\begin{equation}
\psi_{\sf 3}(x,y,t)=e^{ik_{y}y}\sum^{l=+\infty}_{l=-\infty}\left(a_{l}^{3}\varphi^{+}_{l}+b_{l}^{3}\varphi^{-}_{l}\right)e^{-iv_{F}(\epsilon+ l\varpi)t}
\end{equation}
where $\varphi^{\pm}_{l}$ are given by
\begin{eqnarray}
\varphi^{\pm}_{l}&=&\left(%
\begin{array}{c}
 \sqrt{\frac{\epsilon_{l,n}\pm \mu}{\epsilon_{l,n}}}D_{\left[(\epsilon_{l,n}l_{B})^{2}-(\mu l_{B})^{2}\right]/2-1}
 \left[\pm \sqrt{2}\left(\frac{1}{l_{B}}(x-x_{3})+k_{y}l_{B}\right)\right] \\
  \pm i\frac{\sqrt{2/l_{B}^{2}}}{\sqrt{\epsilon_{l,n}(\epsilon_{l,n}\pm \mu)}}
  D_{\left[(\epsilon_{l,n}l_{B})^{2}-(\mu l_{B})^{2}\right]/2}
  \left[\pm \sqrt{2}\left(\frac{1}{l_{B}}(x-x_{3})+k_{y}l_{B}\right)\right] \\
\end{array}%
\right)
\end{eqnarray}

In the forthcoming analysis, we will see how the obtained results so far can be applied to deal with different issues. More precisely, we will
focus on the transmission probability for different channels.

\section{Transmission probability}

Based on different considerations, we study interesting features of our system
in terms of the corresponding transmission probability. Before doing so,
let us simplify our writing using the following shorthand notation
\begin{eqnarray}
&& A_{l,n}^{\pm}=\sqrt{\frac{\epsilon_{l,n}\pm
\mu}{\epsilon_{l,n}}}\\
&&
B_{l,n}^{\pm}=\frac{\sqrt{2/l_{B}^{2}}}{\sqrt{\epsilon_{l,n}(\epsilon_{l,n}\pm
\mu)}}\\
&&\eta_{1,l}^{\pm}=D_{\left[(\epsilon_{l,n}l_{B})^{2}-(\mu
l_{B})^{2}\right]/2-1}
 \left[\pm \sqrt{2}\left(k_{y}l_{B}\right)\right]\\
&& \xi_{1,l}^{\pm}= D_{\left[(\epsilon_{l,n}l_{B})^{2}-(\mu
l_{B})^{2}\right]/2}
  \left[\pm \sqrt{2}\left(k_{y}l_{B}\right)\right]\\
&& \eta_{2,l}^{\pm}=D_{\left[(\epsilon_{l,n}l_{B})^{2}-(\mu
l_{B})^{2}\right]/2-1}
 \left[\pm \sqrt{2}\left(\frac{2d_{1}}{l_{B}}+k_{y}l_{B}\right)\right]\\
&& \xi_{2,l}^{\pm}= D_{\left[(\epsilon_{l,n}l_{B})^{2}-(\mu
l_{B})^{2}\right]/2}
  \left[\pm \sqrt{2}\left(\frac{2d_{1}}{l_{B}}+k_{y}l_{B}\right)\right].
\end{eqnarray}

Realizing that $\{e^{imv_{F}\varpi t}\}$ are orthogonal, we obtain
set of simultaneous equations emanating from the boundary conditions at $x=-d_{2}$
\begin{eqnarray}
&& a_{m}^{1}+b_{m}^{1}=\sum^{l=\infty}_{l=-\infty} \left(a^{2}_{l}
+b^{2}_{l}\right)
J_{m-l}\left(\frac{u_{2}}{\varpi}\right)\lb{feq01}\\
&& a_{m}^{1}z_{m}^{1}-b_{m}^{1}\frac{1}{z_{m}^{1}}=
\sum^{l=\infty}_{l=-\infty}
\left(a^{2}_{l}z_{l}^{2}-b^{2}_{l}\frac{1}{z_{l}^{2}}\right)
J_{m-l}\left(\frac{u_{2}}{\varpi}\right)
\end{eqnarray}
similarly at $x=-d_{1}$
\begin{eqnarray}
&& a^{3}_{m}A^{+}_{m,n}\eta_{1,m}^{+}
+b^{3}_{m}A^{-}_{m,n}\eta_{1,m}^{-} =\sum^{l=\infty}_{l=-\infty}
\left(a^{2}_{l}e^{ik_{l}^{2}(d_{2}-d_{1})}
+b^{2}_{l}e^{-ik_{l}^{2}(d_{2}-d_{1})}\right)
J_{m-l}\left(\frac{u_{2}}{\varpi}\right)\\
&&
a^{3}_{m}iB^{+}_{m,n}\xi_{1,m}^{+}-b^{3}_{m}iB^{-}_{m,n}\xi_{1,m}^{-}=\sum^{l=\infty}_{l=-\infty}
\left(a^{2}_{l}z_{l}^{2}e^{ik_{l}^{2}(d_{2}-d_{1})}-b^{2}_{l}\frac{1}{z_{l}^{2}}e^{-ik_{l}^{2}(d_{2}-d_{1})}\right)
J_{m-l}\left(\frac{u_{2}}{\varpi}\right)
\end{eqnarray}
and at $x=d_{1}$
\begin{eqnarray}
&& a^{3}_{m}A^{+}_{m,n}\eta_{2,m}^{+}
+b^{3}_{m}A^{-}_{m,n}\eta_{2,m}^{-} =\sum^{l=\infty}_{l=-\infty}
\left(a^{4}_{l} +b^{4}_{l}\right)
J_{m-l}\left(\frac{u_{4}}{\varpi}\right)e^{-i(m-l)\delta}\\
&&
a^{3}_{m}iB^{+}_{m,n}\xi_{2,m}^{+}-b^{3}_{m}iB^{-}_{m,n}\xi_{2,m}^{-}=\sum^{l=\infty}_{l=-\infty}
\left(a^{4}_{l}z_{l}^{4}-b^{4}_{l}\frac{1}{z_{l}^{4}}\right)
J_{m-l}\left(\frac{u_{4}}{\varpi}\right)e^{-i(m-l)\delta}.
\end{eqnarray}
However, at $x=d_{2}$ we have
\begin{eqnarray}
&& a_{m}^{5}+b_{m}^{5}=\sum^{l=\infty}_{l=-\infty}
\left(a^{4}_{l}e^{ik_{l}^{4}(d_{2}-d_{1})}
+b^{4}_{l}e^{-ik_{l}^{4}(d_{2}-d_{1})}\right)
J_{m-l}\left(\frac{u_{4}}{\varpi}\right)e^{-i(m-l)\delta}\\
&&a_{m}^{5}z_{m}^{5}-b_{m}^{5}\frac{1}{z_{m}^{5}} =
\sum^{l=\infty}_{l=-\infty}
\left(a^{4}_{l}z_{l}^{4}e^{ik_{l}^{4}(d_{2}-d_{1})}-b^{4}_{l}\frac{1}{z_{l}^{4}}e^{-ik_{l}^{4}(d_{2}-d_{1})}\right)
J_{m-l}\left(\frac{u_{4}}{\varpi}\right)e^{-i(m-l)\delta}.
\lb{seq01}
\end{eqnarray}

As Dirac electrons pass through a region subject to
time-harmonic potentials, transitions from the central band to
sidebands (channels) at energies $\epsilon\pm m\varpi$ $(m = 0, 1,
2, \cdots)$ occur as electrons ex- change energy quanta with the
oscillating field. It should be noted that
(\ref{feq01}-\ref{seq01})  can be written in a compact form as
\begin{eqnarray}
\left(%
\begin{array}{c}
  \Xi_{1} \\
  \Xi_{1}^{'} \\
\end{array}%
\right)=\left(%
\begin{array}{cc}
 { \mathbb M11} &{\mathbb M12} \\
 {\mathbb M21} &{ \mathbb M22} \\
\end{array}%
\right)\left(%
\begin{array}{c}
  \Xi_{5} \\
  \Xi_{5}^{'}\\
\end{array}%
\right)={\mathbb M}\left(%
\begin{array}{c}
  \Xi_{5} \\
 \Xi_{5}^{'} \\
\end{array}%
\right) \lb{feqi}
\end{eqnarray}
where the total transfer matrix ${\mathbb M}={\mathbb
M(1,2)} \cdot {\mathbb M(2,3)} \cdot {\mathbb M(3,4)} \cdot {\mathbb M(4,5)}$ and
${\mathbb M(j,j+1)}$ are transfer matrices that couple the wave
function in the $j$-th region to that in the $(j + 1)$-th one.
These are explicitly defined  by
\begin{eqnarray}
{\mathbb M(1,2)}=\left(%
\begin{array}{cc}
  {\mathbb I}& {\mathbb I} \\
{\mathbb N^{+}} &{\mathbb N^{-}} \\
\end{array}%
\right)^{-1}
\left(%
\begin{array}{cc}
  {\mathbb C} & {\mathbb C} \\
 {\mathbb G^{+}} & {\mathbb G^{-}} \\
\end{array}%
\right)
\end{eqnarray}

\begin{eqnarray}
{\mathbb M(2,3)}=\left(%
\begin{array}{cc}
  {\mathbb Y_{1}^{+}}& {\mathbb Y_{1}^{-}} \\
{\mathbb Y_{2}^{+}} &{\mathbb Y_{2}^{-}} \\
\end{array}%
\right)^{-1}
\left(%
\begin{array}{cc}
  {\mathbb Q_{1}^{+}} & {\mathbb Q_{1}^{-}} \\
 {\mathbb F_{1}^{+}} & {\mathbb F_{1}^{-}} \\
\end{array}%
\right)
\end{eqnarray}

\begin{eqnarray}
{\mathbb M(3,4)}=\left(%
\begin{array}{cc}
  {\mathbb Q_{2}^{+}} & {\mathbb Q_{2}^{-}} \\
  {\mathbb F_{2}^{+}} & {\mathbb F_{2}^{-}} \\
\end{array}%
\right)^{-1}
\left(%
\begin{array}{cc}
  {\mathbb D_{1}}& {\mathbb D_{1}} \\
{\mathbb D_{2}^{+}} &{\mathbb D_{2}^{-}} \\
\end{array}%
\right)
\end{eqnarray}
\begin{eqnarray}
{\mathbb M(4,5)}=\left(%
\begin{array}{cc}
  {\mathbb K_{1}^{+}} & {\mathbb K_{1}^{-}} \\
  {\mathbb K_{2}^{+}} & {\mathbb K_{2}^{-}} \\
\end{array}%
\right)^{-1}
\left(%
\begin{array}{cc}
  {\mathbb I}& {\mathbb I} \\
{\mathbb E^{+}} &{\mathbb E^{-}} \\
\end{array}%
\right)
\end{eqnarray}
whose  matrix elements  are expressed as
\begin{eqnarray}
&&\left({\mathbb
N^{\pm}}\right)_{m,l}=\pm\left(z_{m}^{1}\right)^{\pm
1}\delta_{m,l}\\
&&
\left({\mathbb
C}\right)_{m,l}=J_{m-l}\left(\frac{u_{2}}{\varpi}\right)\\
&&
\left({\mathbb G^{\pm}}\right)_{m,l}=\pm(z_{l}^{1})^{\pm
1}J_{m-l}\left(\frac{u_{2}}{\varpi}\right)\\
&&
\left({\mathbb Y_{1}^{\pm}}\right)_{m,l}=e^{\pm
ik_{l}^{2}(d_{2}-d_{1})}J_{m-l}\left(\frac{u_{2}}{\varpi}\right)\\
&&
\left({\mathbb Y_{2}^{\pm}}\right)_{m,l}=\pm(z_{l}^{2})^{\pm
1}e^{\pm
ik_{l}^{2}(d_{2}-d_{1})}J_{m-l}\left(\frac{u_{2}}{\varpi}\right)\\
&&
\left({\mathbb
Q_{\tau}^{\pm}}\right)_{m,l}=A_{m,n}^{\pm}\eta_{\tau,m}^{\pm}\delta_{m,l}\\
&&
\left({\mathbb F_{\tau}^{\pm}}\right)_{m,l}=\pm
iB_{m,n}\xi_{\tau,m}^{\pm}\delta_{m,l}\\
&&
\left({\mathbb
D_{1}}\right)_{m,l}=J_{m-l}\left(\frac{u_{4}}{\varpi}\right)e^{\left(-i(m-l)\delta\right)}\\
&&
\left({\mathbb D_{2}^{\pm}}\right)_{m,l}=\pm(z_{l}^{4})^{\pm
1}J_{m-l}\left(\frac{u_{4}}{\varpi}\right)e^{\left(-i(m-l)\delta\right)}\\
&&
\left({\mathbb K_{1}^{\pm}}\right)_{m,l}=e^{\pm
ik_{l}^{4}(d_{2}-d_{1})}J_{m-l}\left(\frac{u_{4}}{\varpi}\right)e^{\left(-i(m-l)\delta\right)}\\
&& \left({\mathbb
 K_{2}^{\pm}}\right)_{m,l}=\pm(z_{l}^{4})^{\pm
1}e^{\pm
ik_{l}^{4}(d_{2}-d_{1})}J_{m-l}\left(\frac{u_{4}}{\varpi}\right)e^{\left(-i(m-l)\delta\right)}\\
&& \left({\mathbb
E^{\pm}}\right)_{m,l}=\pm\left(z_{m}^{5}\right)^{\pm
1}\delta_{m,l}
\end{eqnarray}
and the unit matrix is denoted by  ${\mathbb I}$. We assume an
electron propagating from left to right with quasienergy
$\epsilon$. Then, $\tau \in\{1,2\}$,
$\Xi_{1}=\{a_{m}^{1}\}=\{\delta_{m,0}\}$ and
$\Xi_{5}^{'}=\{b_{m}^{5}\}$ is the null vector, whereas
$\Xi_{5}=\{a_{m}^{5}\}=\{t_{m}\}$ and $\Xi_{1}^{'}=\{b_{m}^{1}\}
=\{r_{m}\}$ are vectors associated with transmitted waves and
reflected waves, respectively. From the above considerations, one
can easily obtain the relation
\begin{eqnarray}
\Xi_{5}=\left({ \mathbb M11}\right)^{-1} \cdot \Xi_{1}
\end{eqnarray}
which is equivalent to the explicit form
\begin{eqnarray}
\left(%
\begin{array}{cc}
 t_{-N} \\
. \\
. \\
 t_{-1}\\
 t_{0} \\
t_{1} \\
 .\\
 .\\
 t_{N}\\
\end{array}%
\right)=\left({ \mathbb M11}\right)^{-1}
\left(%
\begin{array}{cc}
 0 \\
0 \\
0 \\
 0\\
 1 \\
0 \\
0 \\
 0\\
 0\\
\end{array}%
\right)
\end{eqnarray}
The minimum number $N$ of sidebands that need to be taken is
determined by the strength of the oscillating potential,
$N>\mbox{max}\left(\frac{u_{2}}{\varpi}, \frac{u_{4}}{\varpi}\right)$
\cite{ahsan}. Then the infinite series for the transmission $T$ can be truncated
considering only a finite number of terms starting from $-N$ up to $N$.
Furthermore, analytical results are obtained if we pick up small
values of $\alpha_{2}=\frac{u_{2}}{\varpi}$,
$\alpha_{4}=\frac{u_{4}}{\varpi}$ and include only the first two
sidebands at energies $\epsilon\pm m \varpi$ along with the
central band at energy $\epsilon$. This gives
\begin{equation}
t_{-N+k}={ \mathbb M^{'}}\left[k+1, N+1\right)]
\end{equation}
where $k=0, 1, 2, \cdots, 2N$ and ${ \mathbb M^{'}}$ denotes the inverse matrix
$\left({ \mathbb M11}\right)^{-1}$.

Using the reflected $J_{\sf {ref}}$ and transmitted $J_{\sf {tra}}$ currents
to write the reflection and transmission coefficients
$R_{l}$ and $T_{l}$ as 
\begin{equation}
  T_{l}=\frac{ |J_{{\sf {tra}},l}|}{| J_{{\sf {inc}},0}|},\qquad R_{l}=\frac{|J_{{\sf {ref}},l}|}{ |J_{\sf {inc,0}}|}
\end{equation}
where $T_{l}$ is the transmission coefficient describing the
scattering of an electron with incident quasienergy $\epsilon$ in
the region 1 into the sideband with quasienergy $\epsilon+l\varpi$
in the region 5.  Thus, the rank of the transfer matrix $({\mathbb
M})$ increases with the amplitude of the time-oscillating
potential. The total transmission coefficient for quasienergy
$\epsilon$ is
\begin{equation}
T=\sum_{l=-\infty}^{l=+\infty}T_{l}.
\end{equation}
The electrical current density $J$ corresponding to our system can be derived to be 
\begin{equation}
J= v_{F}\psi^{\dagger}\sigma _{x}\psi
\end{equation}
which  explicitly reads as
\begin{eqnarray}
&& J_{{\sf {inc}}, 0}= v_{F} \left(z_{0}^{1}+\left(z_{0}^{1}\right)^{\ast}\right)\\
&&
  J_{{\sf {ref}}, l}= v_{F} \left(b_{l}^{1}\right)^{\ast}b_{l}^{1}\left(z_{l}^{1}+\left(z_{l}^{1}\right)^{\ast} \right)\\
&&
 J_{{\sf {tra}}, l}= v_{F} \left(a_{l}^{5}\right)^{\ast}a_{l}^{5}\left(z_{l}^{5}+\left(z_{l}^{5}\right)^{\ast} \right).
\end{eqnarray}
The transmission coefficient for the sideband, $T_{l}$,
is real and corresponds to propagating waves. It can be written as
\begin{equation}
  T_{l}= \frac{s_{l}^{5} k^{5}_{l}}{s_{0}^{1} k^{1}_{0}}\frac{\left[ \left(k^{1}_{0} \right)^{2}+ \left(k_{y}-
  \frac{d_{1}}{l_{B}^{2}}\right)^{2}\right]^{\frac{1}{2}}}{\left[\left(k^{5}_{l} \right)^{2}+
  \left(k_{y}+\frac{d_{1}}{l_{B}^{2}} \right)^{2}\right]^{\frac{1}{2}}}\mid t_{l}^{5}\mid^{2}.
\end{equation}
Now using the energy conservation to simplify $T_{l}$ to
\begin{equation}
  T_{l}= \frac{k^{5}_{l}}{k^{1}_{0}}\left( 1-\frac{s_{l}^{5} l\varpi}{\sqrt{\left(k^{5}_{l} \right)^{2}+
  \left(k_{y}+\frac{d_{1}}{l_{B}^{2}} \right)^{2}}}  \right)
  \mid t_{l}^{5}\mid^{2}.
\end{equation}

To explore the above results and go deeply in order to underline our system behavior,
we will pass to the numerical analysis. For this, we will focus only on few channels
and choose different configurations of the physical parameters.

\section{Discussions} 

We discuss the numerical results for both the
reflection and transmission coefficients, which are shown in Figures
2, 3, 4, 5, 6, 7, 8, 9 for different values of the parameters
($\epsilon$, $v_{2}$, $v_{3}$, $d_{1}$, $d_{2}$, $\alpha_{2}$,
$\alpha_{4}$, $\delta$, $\mu$).
To start with we point out the efficiency and accuracy of our computational method and compare our results with those
reported in the literature.
As a matter of fact,  Figure \ref{fig1920}(a) and
Figure \ref{fig1920}(b) reproduce exactly the results
obtained in \cite{martino} for single barrier and \cite{mekkaoui} for double barrier, respectively, with the proper choice of parameters.
Note that reference \cite{martino} was the first to treat the
confinement of Dirac fermions by an inhomogeneous magnetic field.
These polar graphs show the transmission as a function of the
incidence angle, the outermost circle corresponds to full
transmission, $T_{0} = 1$, while the origin of this plot
represents zero transmission, i.e. total reflection.
For energies satisfying
the  condition $(\epsilon+l\varpi) l_{B}\leq \frac{d_{1}}{l_{B}}$, we obtain total reflection \cite{martino,mekkaoui}. This is
equivalent to the condition on the incidence angle $\phi < \phi_c$ where $\phi_c$ is
the critical angle given by
\begin{equation}
    \phi_c = \sin^{-1}\left (1-\frac{2d}{(\epsilon+l\varpi) l^2_B}
    \right)\lb{eqch}
\end{equation}
which is analogous to the case of light propagation from a refringent medium to a less refringent one.

\begin{figure}[h!]
\centering
\includegraphics[width=6cm, height=9cm]{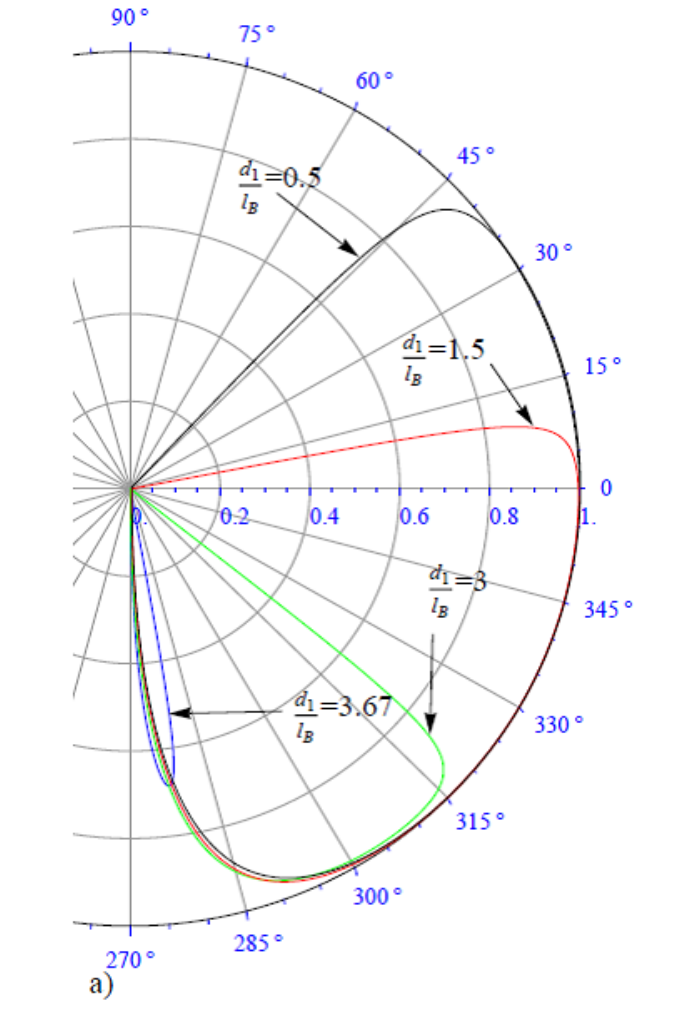}\ \ \ \
\includegraphics[width=6cm, height=9cm]{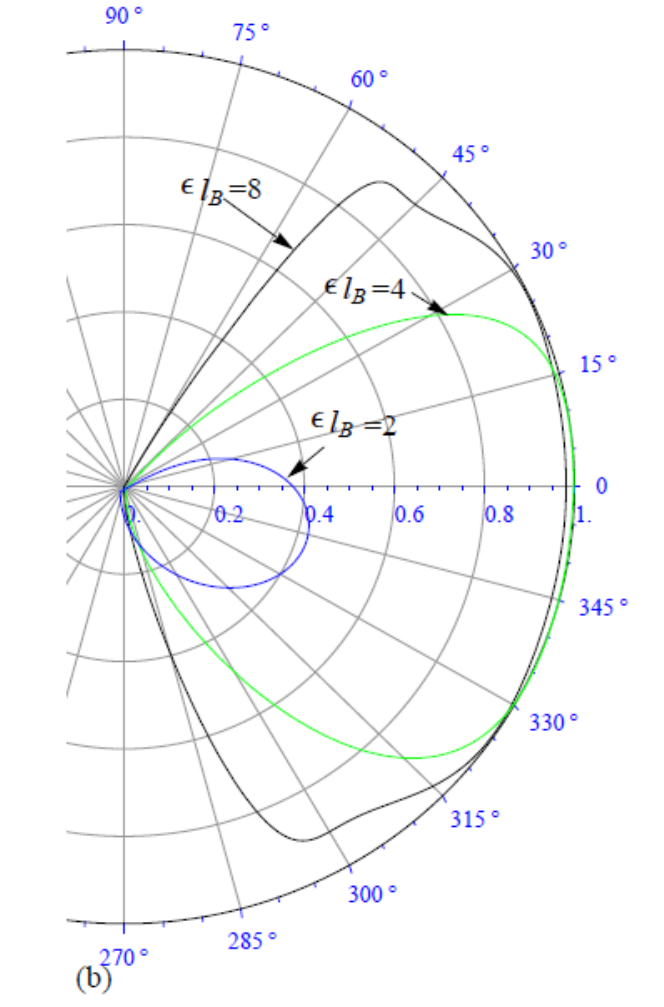}\\
 \caption{\sf{(Color online) Polar plot showing transmission probability (transmission $T_{0}$ ($l$=0)) as a function of angle $\phi_{0}^{1}$
 for different values of the parameters.
(a): $\frac{d_{1}}{l_{B}}=\{0.5, 1.5, 3, 3.67\}$,  $\epsilon
 l_{B}=3.7$,
$\alpha_{2}=\alpha_{4}=0 $, $\delta=0$,
$\frac{d_{2}}{l_{B}}=\frac{d_{1}}{l_{B}}$, $v_{2} l_{B}=v_{4}
l_{B}=0$,
 $\mu l_{B}=0$ and $v_{3}l_{B}=0$. (b):   $\epsilon
 l_{B}=\{0.6, 2, 4, 8\}$, $\frac{d_{1}}{l_{B}}=0.5$,
$\alpha_{2}=\alpha_{4}=0 $, $\delta=0$, $\frac{d_{2}}{l_{B}}=0.6$,
$v_{2} l_{B}=v_{4} l_{B}=0.5$,
 $\mu l_{B}=1$ and $v_{3}l_{B}=0.4$ .}}\lb{fig1920}
\end{figure}

After a satisfactory confirmation that our numerical approach reproduced
published results, we plot the transmission versus the phase shift $\delta$
in the presence temporal barrier oscillations for the cases 
$\alpha=\alpha_{2}=\alpha_{4}$ and $\alpha=\alpha_{2}=2\alpha_{4}$.
Figure \ref{Figch12} illustrates
the variation of the transmission coefficient $T_0$ 
as a function of the phase shift $\delta$ for various harmonic amplitudes $\alpha$
($0\leq\alpha\leq0.99$).
The first point to emphasize is that both plots in Figure \ref{Figch12} are
periodic with period $2\pi$, which is obvious. For $\alpha = 0$, the
transmission does not depend on the phase shift since the
vibration amplitude has been set to zero.
For $\alpha=\alpha_{2}=\alpha_{4}$, the transmission varies sinusoidally  between 0 and 1,
the maximum for different plots does not change for various values of $\al$ (Figure \ref{Figch12}(a)).
However, when the harmonic oscillation amplitudes are not equal ($\alpha_2\neq \alpha_4$), we observe that
there is a remarkable change
in the evolution of $T_0$ versus $\delta$. We illustrate this situation by selecting $\alpha_2 = 2 \alpha_4$
then one can  see that as long as such difference increases we observe a drastic change in the transmission
from full transmission (total transmission) to zero transmission (total reflection), see Figure \ref{Figch12}(b).

\begin{figure}[h!]
\centering
\includegraphics[width=8
cm,height=5cm]{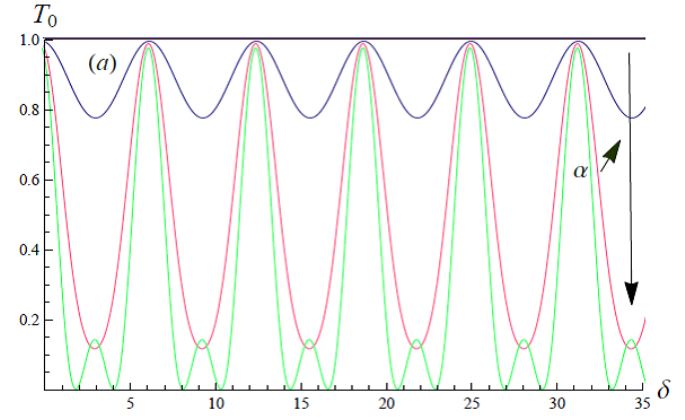} \ \ \ \
\includegraphics[width=8
cm,height=5cm]{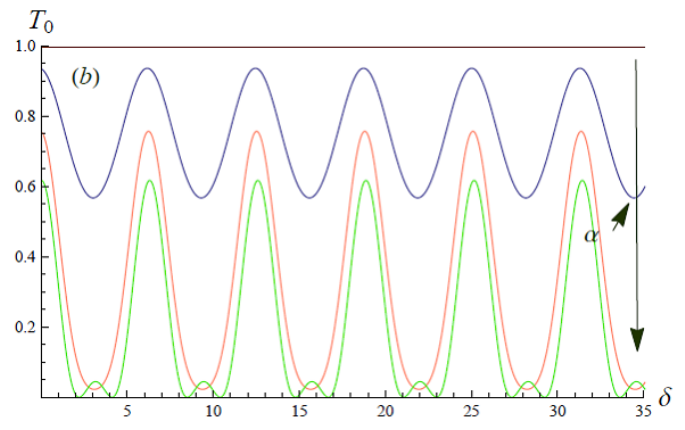}
 \caption{\sf{(Color online) The transmission coefficient $T_{0}$ as  function of phase shift
$\delta$ through graphene  double barriers  for fixed
values $\epsilon l_{B}$, $k_{y}l_{B}$, $\frac{d_{1}}{l_{B}}$,
$\frac{d_{2}}{l_{B}}$, $\mu l_{B}$, $v_{3} l_{B}$, $\varpi l_{B}$
and $v_{2}l_{B}=v_{4}l_{B}$ but for different values of $\alpha$.
We used $\epsilon l_{B}=25$,
 $k_{y}l_{B}=2$,  $\frac{d_{1}}{l_{B}}=0.3$,  $\frac{d_{2}}{l_{B}}=1.35$,  $\mu l_{B}=4$, $v_{3} l_{B}=4$,
  $\varpi l_{B}=2$ and  $v_{2}l_{B}=v_{4}l_{B}=6$
  and $\alpha$ varies from $0$
to $0.99$. (a): $\alpha=\alpha_{2}=\alpha_{4}$, (b):
$\alpha=\alpha_{2}=2\alpha_{4}$  }}\lb{Figch12}
\end{figure}

Let us now demonstrate through Figure \ref{fig5} how the first sideband transmissions $T_1$ and
$T_{-1}$ vary as function of  the phase shift $\delta$. It is clearly seen that
the central transmission $T_0$ behaves sinusoidally but at some value of $\alpha$
we observe that $T_0$ changes its behavior and becomes sharply peaked.
$T_1$ and $T_{-1}$ show also
sinusoidal behaviors with 
non-symmetric double humps in regions where $T_0$ is suppressed.
However, one can see that there is a symmetry between the two double humps of the
transmissions $T_1$ and $T_{-1}$.

\begin{figure}[H]
\centering
\includegraphics[width=8cm, height=5cm]{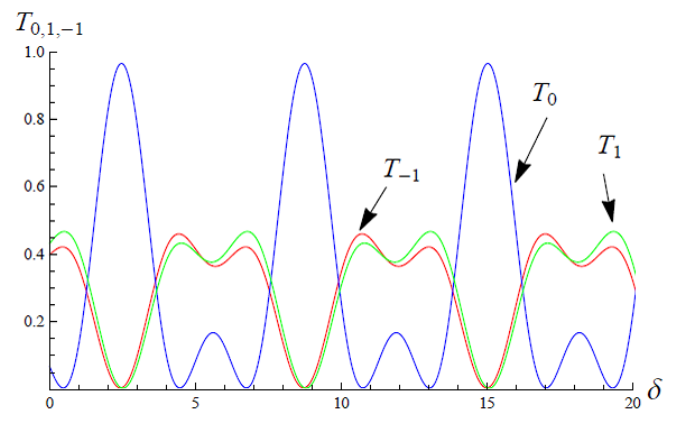}\\
 \caption{\sf{(Color online)
Graphs depicting the transmission probabilities  as  function of
phase $\delta$ for  graphene double barriers with $\varpi
l_{B}= 8$,
  $\alpha_{2}=\alpha_{4}= 0.8 $, $\epsilon l_{B}=25$,
 $k_{y}l_{B}=2$,  $\frac{d_{1}}{l_{B}}=0.3$,  $\frac{d_{2}}{l_{B}}=1.35$,
  $\mu l_{B}=4$, $v_{3} l_{B}=4$ and
  $v_{2}l_{B}=v_{4}l_{B}=6$.}}\lb{fig5}
\end{figure}

\begin{figure}[H]
\centering
\includegraphics[width=8cm, height=5cm]{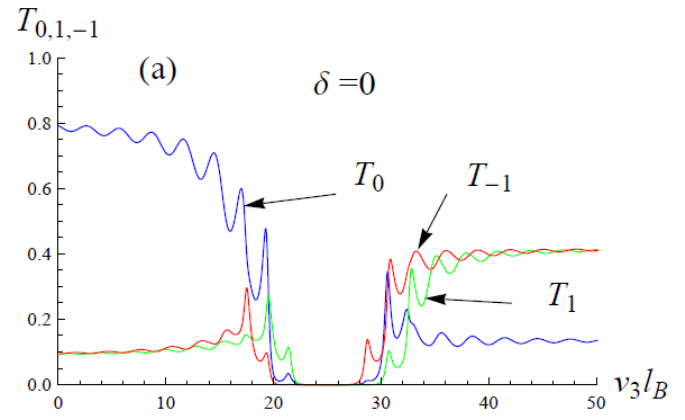}\ \ \ \
\includegraphics[width=8cm, height=5cm]{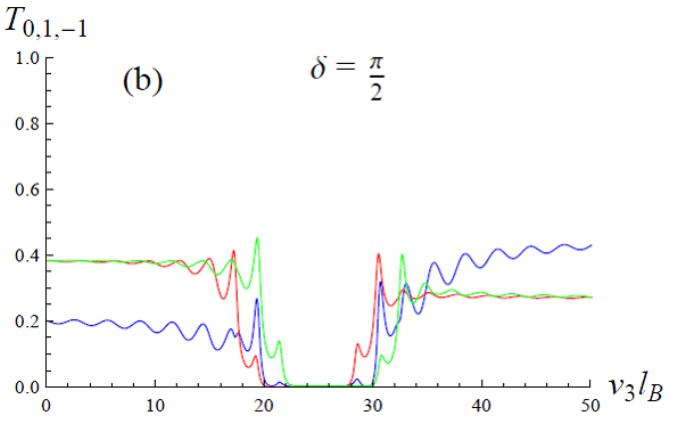}\\
\includegraphics[width=8cm, height=5cm]{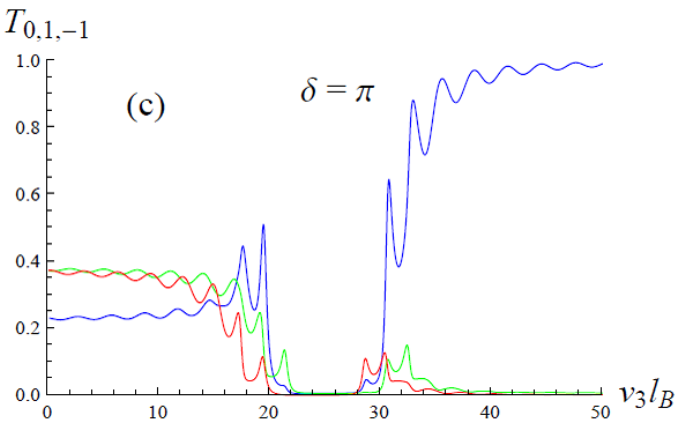}\ \ \ \
\includegraphics[width=8cm, height=5cm]{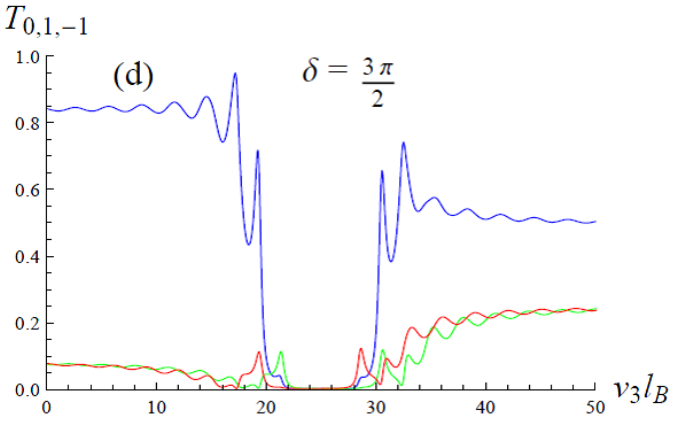}
 \caption{\sf{(Color online) Graphs depicting the transmission probabilities as  function of
potential  $v_{3}l_{B}$ for the monolayer graphene barriers with
$\frac{d_{2}}{l_{B}}=1.5$, $\frac{d_{1}}{l_{B}}=0.5$, $v_{2}
l_{B}=v_{4} l_{B}=4$, $\epsilon l_{B}=25$,
 $k_{y}l_{B}=2$, $\mu l_{B}=4$, $\varpi l_{B}=2$, $\alpha_{2}=\alpha_{4}=0.5 $ and $\delta=\{0, \frac{\pi}{2}, \pi,
 \frac{3\pi}{2}\}$.
 $T_{0}$ (color blue), $T_{-1}$ (color red) and $T_{1}$ (color green).}}\lb{fig6}
\end{figure}

Now we will study how the three bands: central and two lateral ones, vary
depending on the phase difference of the oscillating potentials in the
intermediate region (Figure \ref{fig6}).
For different phase shifts, the  transmissions of side bands
$T_{1}$ and $T_{-1}$ are dominant either before or after the
bowl centered region in the propagation energy  $\epsilon l_{B}$. In the
vicinity of this bowl, one of the two transmissions is more
symmetrical with respect to a vertical axis
passing through the energy $\epsilon l_{B}$. The degree of dominance of
the transmissions $T_{1}$ and $T_{-1}$  is less pronounced in the case of
advanced phase quadrature $\delta = \frac{\pi}{2}$ (Figure \ref{fig6}(b)).
But they are more dominant in the case of $\delta = 0$ and $\delta=
\frac{3\pi}{2}$ where the potential $v_{3} l_{B}$ is greater than
the propagation energy $\epsilon l_{B}$ and are less dominant
when  $v_{3} l_{B}$ is less than the propagation energy $\epsilon
l_{B}$. The behavior of the transmission side bands differs in the
case of opposite phase shift $\delta = \pi$. On the other hand,
the transmission of the central band, for different phase shifts,
has also a dominance of either side of the high potential  $v_{3}
l_{B}$ than the propagation energy $\epsilon l_{B}$ or the other
side where the potential  $v_{3} l_{B}$ is small than the same
energy.

\begin{figure}[H]
\centering
\includegraphics[width=8cm, height=5cm]{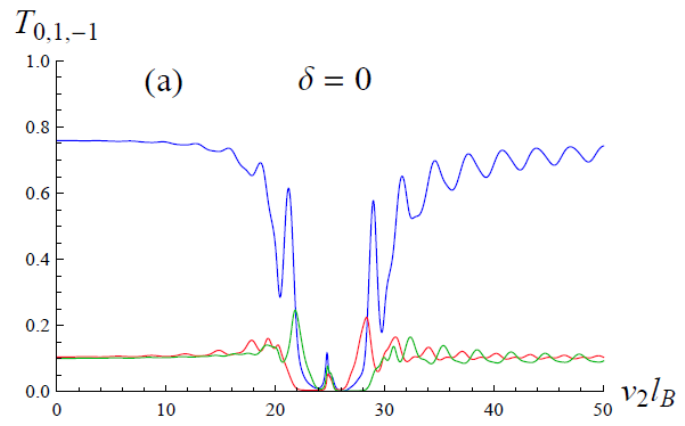}\ \ \ \
\includegraphics[width=8cm, height=5cm]{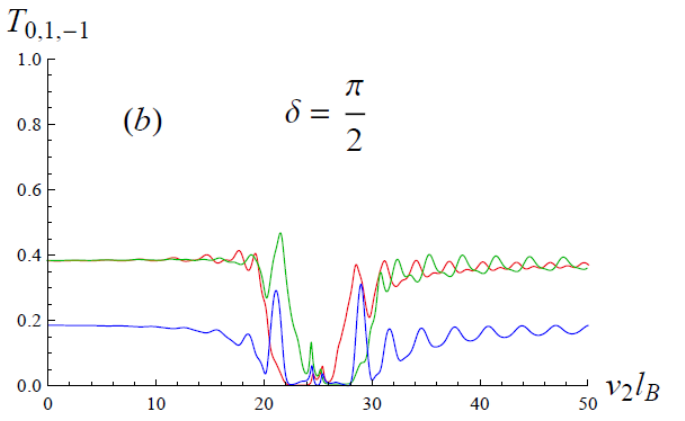}\\
\includegraphics[width=8cm, height=5cm]{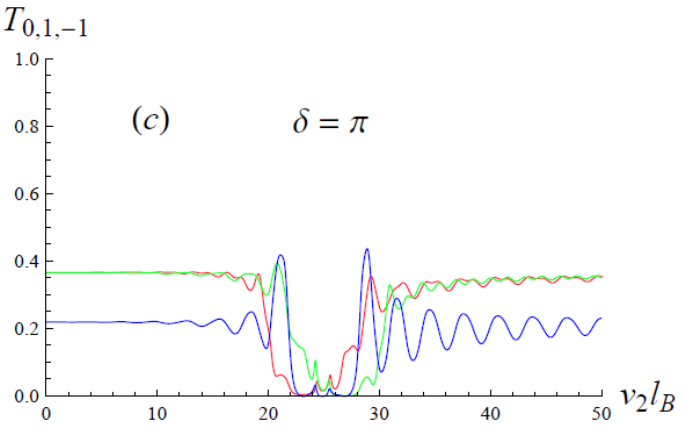}\ \ \ \
\includegraphics[width=8cm, height=5cm]{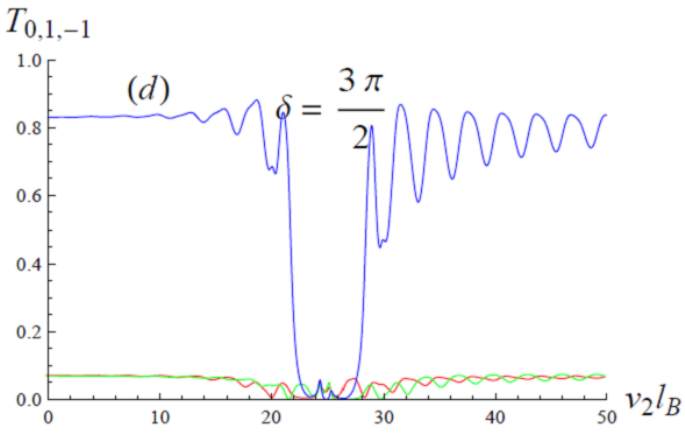}
 \caption{\sf{ (Color online)
Graphs depicting the transmission probabilities  as  function of
 potential  $v_{2}l_{B}$ for monolayer graphene barriers
with
  $\alpha_{2}=\alpha_{4}= 0.5 $, $\varpi l_{B}=2$, $\epsilon l_{B}=25$,
 $k_{y}l_{B}=2$,  $\frac{d_{1}}{l_{B}}=0.5$,  $\frac{d_{2}}{l_{B}}=1.6$,  $\mu l_{B}=4$, $v_{3} l_{B}=4$ and  $\delta=\{0, \frac{\pi}{2}, \pi,
 \frac{3\pi}{2}\}$.
 $T_{0}$ (color blue), $T_{-1}$ (color red) and $T_{1}$ (color green).}}\lb{fig8}
\end{figure}

{Figure \ref{fig8} is similar to Figure \ref{fig6}, 
the only main difference to be noted is that there is
 presence of peaks in the bowl centered around the value
$v_{2}=\epsilon$ and  $T_{0,-1,1}$ heights are in the same order either before or after the bowl. These peaks are due to the resonances between the
 bound states existing  in  both sides of the regions subject to
the potential $v_{3}$. This behavior is normal if we keep in mind that our double barrier is composed of two successive
squares with the same potential $v_2$  and width
$(d_2-d_1)$ separated by the width $2d_1$ corresponding to the
intermediate region. 

{Figure \ref{fig08} shows the transmission
probability as a function of the incident energy of electrons for
$v_{2} l_{B}=25$, $k_{y}l_{B}=2$,  $\frac{d_{1}}{l_{B}}=0.5$,
$\frac{d_{2}}{l_{B}}=1.5$,  $\mu l_{B}=4$, $v_{3} l_{B}=4$ and
different amplitudes of the oscillating barrier without shift $\delta=0$.
Resonant peaks are narrow and could have  important
applications in high-speed devices based on graphene as has been
suggested previously \cite{ahsan}. The evolution of the central
and two lateral transmission bands depend on the width of the
double barrier potential over time accompanied by a magnetic field
(Figures \ref{fig08}(a), \ref{fig08}(b)).}

\begin{figure}[H]
\centering
\includegraphics[width=8cm, height=5cm]{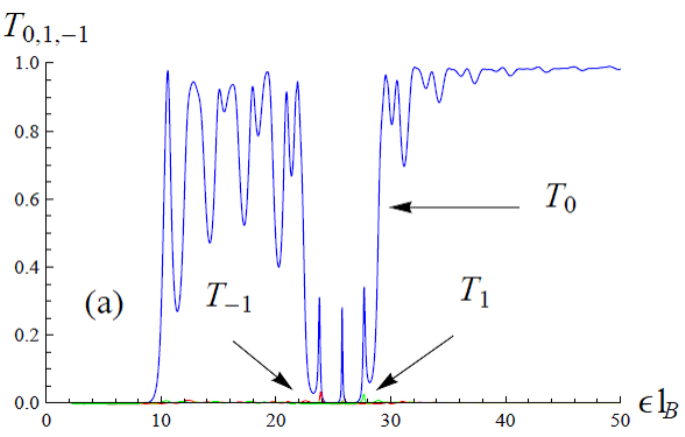}\ \ \ \
\includegraphics[width=8cm, height=5cm]{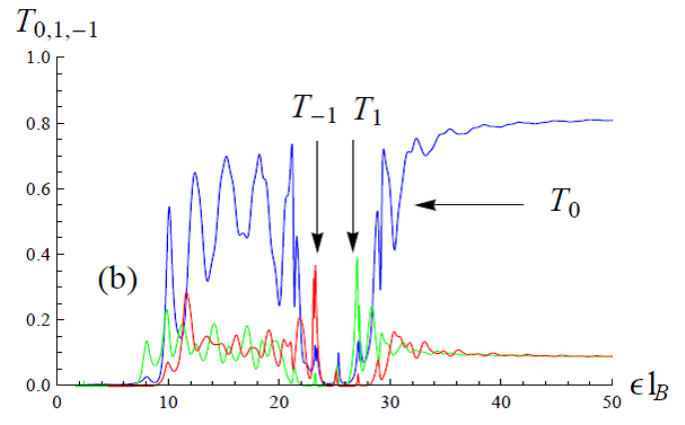}\\
\caption{\sf{ (Color online) Graphs depicting the transmission
probabilities  as  function of energy $\epsilon l_{B}$ for the
monolayer graphene barriers with
  $\alpha_{2}=\alpha_{4}= \{ 0.08, 0.5\} $, $v_{2} l_{B}=25$,
 $k_{y}l_{B}=2$,  $\frac{d_{1}}{l_{B}}=0.5$,  $\frac{d_{2}}{l_{B}}=1.5$,  $\mu l_{B}=4$, $v_{3} l_{B}=4$, $\varpi
l_{B}=2$ and  $\delta=0$.}}\lb{fig08}
\end{figure}

\begin{figure}[H]
\centering
\includegraphics[width=8cm, height=5cm]{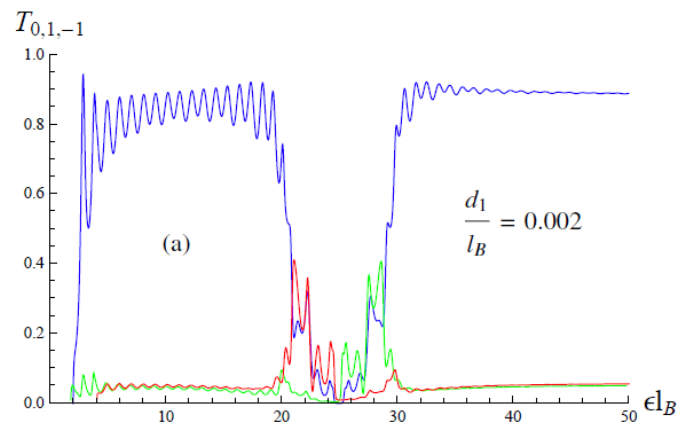}\ \ \ \
\includegraphics[width=8cm, height=5cm]{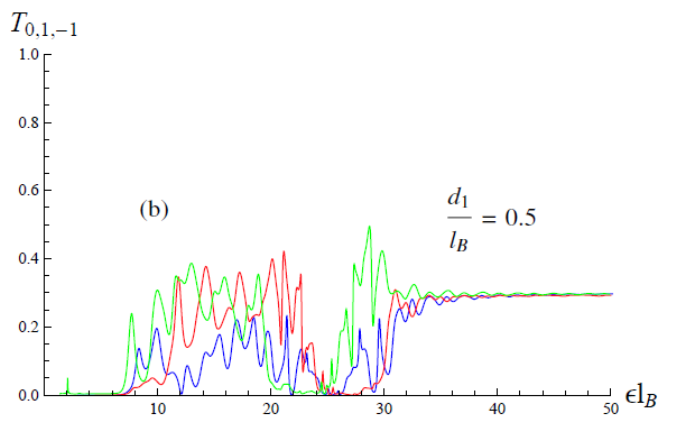}\\
\includegraphics[width=8cm, height=5cm]{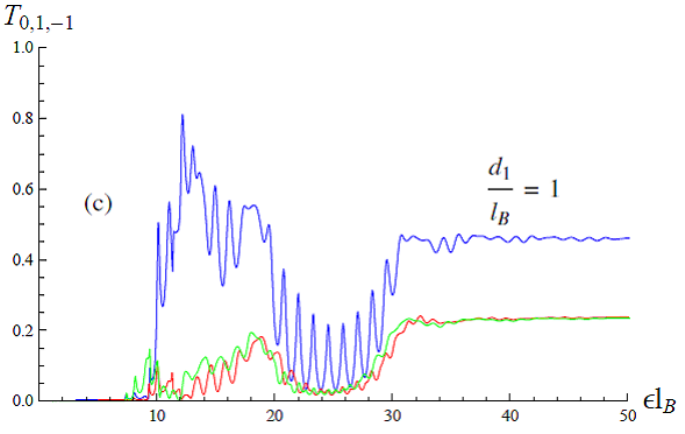}\ \ \ \
\includegraphics[width=8cm, height=5cm]{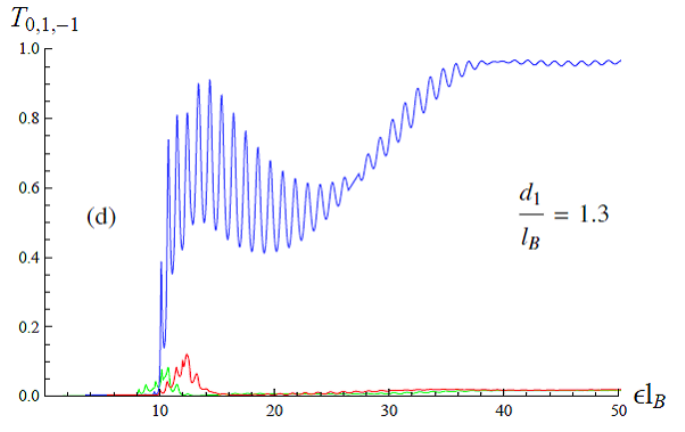}
 \caption{\sf{ (Color online)
Graphs depicting the transmission probabilities  as  function of
 energy  $\epsilon l_{B}$ for monolayer graphene barriers
with
  $\alpha_{2}=\alpha_{4}= 0.99 $, $\frac{d_{1}}{l_{B}}=\{0,0.02,0.5,1,1.3\}$, $v_{2}
  l_{B}=25$, $k_{y}l_{B}=2$, $\frac{d_{2}}{l_{B}}=1.5$, $\mu l_{B}=4$, $v_{3}
  l_{B}=4$, $\varpi l_{B}=2$   and  $\delta=\pi$.
 $T_{0}$ (color blue), $T_{-1}$ (color red) and $T_{1}$ (color green).}}\lb{jell01}
\end{figure}

{Figure \ref{jell01} presents transmission versus the system energy for different widths. Indeed,
we observe that in Figure \ref{jell01}(a) as long as the width is very small
  the central band is
  dominant and therefore the transmission becomes total independently of the
applied potential.
Figure \ref{jell01}(b) is obtained by increasing the width $d$ up to some value, one can see 
the dominance of the two sideband transmissions compared  to central band one. We notice that
these two sideband transmissions are symmetrical with respect to an axis of symmetry located
at double barrier potential $v_2$ of the propagation energy.
After increasing the width $d$, we end up with Figure \ref{jell01}(c), which is similar to the last one but this time
with
dominance of the central band transmission. It is clearly seen that  the total transmission is less than or equal to unity.
In Figure \ref{jell01}(d), the central band transmission
recovers its dominance but evanescence of two sideband transmissions.}



\begin{figure}[H]
\centering
\includegraphics[width=8
cm,height=5cm]{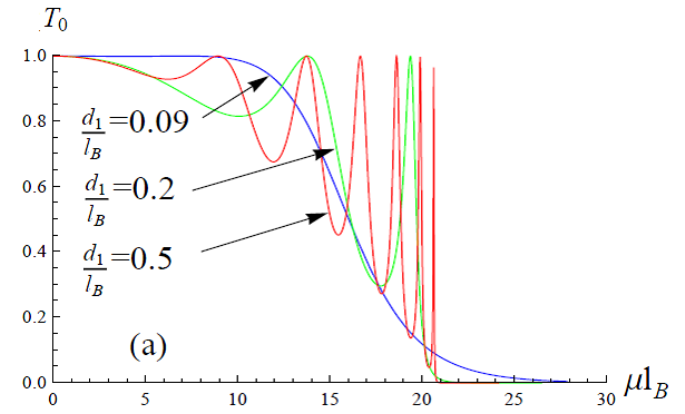}\ \ \
\includegraphics[width=8
cm,height=5cm]{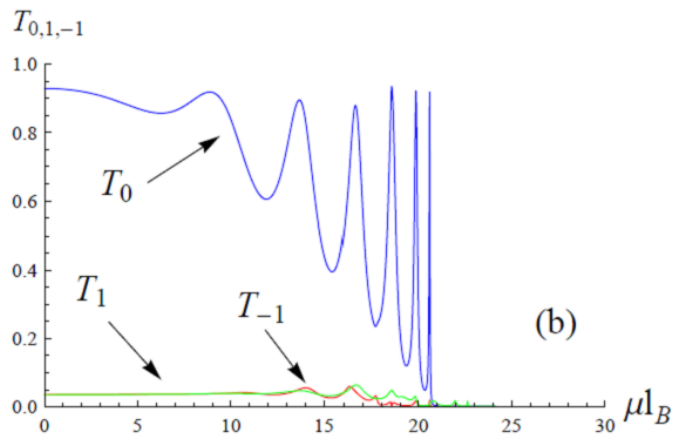}
 \caption{\sf{ (Color online)
Graphs depicting the transmission total $T$ as  function of energy
gap $\mu l_{B}$ for the monolayer graphene barriers with:
$\frac{d_{2}}{l_{B}}=0.7$, $v_{2} l_{B}=v_{4} l_{B}=4$, $\epsilon
l_{B}=25$, $k_{y}l_{B}=2$, $v_{3}l_{B}=4$, $\varpi l_{B}=2$,
$\alpha_{2}=\alpha_{4}=0 $ and $\delta=0$ (a):
$\frac{d_{1}}{l_{B}}=\{0.09, 0.2, 0.5\}$,
$\alpha_{2}=\alpha_{4}=0$ and $\delta=0$. (b):
$\frac{d_{1}}{l_{B}}=0.5$, $\alpha_{2}=\alpha_{4}=0.7$ and
$\delta=\frac{3\pi}{4}$ .}}\lb{fig10}
\end{figure}

 Figure \ref{fig10} is intended to see the influence
of  increasing  the width of the intermediate zone, where there
is a magnetic field, on the dominant transmission central
depending on the mass term $\mu l_{B}$ that in the intermediate
region. The distance $d_2$ remains constant which means that the
widths of regions 2 and 4 decrease if $d_1$ increases. Figure
\ref{fig10}(a) shows that progressively as the distance $d_1$
increases, the central transmission acquires resonances which
clamp by increasing amplitudes whose upper peaks correspond to a
total transmission (maximum). The maximum value of $T_0(\mu)$ is
the unit since $\alpha=\alpha_{2} = \alpha_{4} =  0$.  In Figure
\ref{fig10}(b) for $\alpha \neq 0$ the maximum value of $T_0(\mu)$
decreases at the expense of transmission sidebands
$T_{-1,1}(\mu)$. We note that the sum of the three
transmissions $T_{0,-1,1}(\mu)$ converges whenever towards unity.

\section{Conclusion}

 In this present work, we studied the transmission probability in
graphene through double barriers with periodic potential in time.
The double barrier contains an intermediate region has a magnetic
field with a mass term, but the two temporal harmonic potentials
with different amplitudes and phase shifted are applied one hand
and on the other in both regions restricting the intermediate
region. This panoply of potential makes our studied system rich in
terms of physical states whose energy is doubling quantified by
the pair $(n, l)$ extensively degenerated with a very large number
of modes.

To identify the difficulties posed, we made the problem
by adequate truncation to reduce all modes in three modes one
central and two lateral indexed by $(0,-1,1$). 
We
tried to study the influence of various parameters such as
($\epsilon$, $v_{2}$, $v_{3}$, $d_{1}$, $d_{2}$, $\alpha_{2}$,
$\alpha_{4}$, $\delta$, $\mu$) on the transmission  probability
and highlight some properties of the system under consideration. The
critical angle (see (\ref{eqch})), at which
total reflection sets in,  
showed in an
efficient manner the analogy between the propagation of Dirac
fermions in our system and the propagation of the light of a more
refractive homogeneous isotropic transparent medium a less refractive one.
This built an interesting bridge between two areas of
physics such matter and light.}

 The
transmission probability $T_0(\delta)$ is obtained to be  harmonic with frequency
proportional to $\varpi$ of time dependent amplitude $\alpha$. We observed that
as long as  the amplitude $\alpha$ of time-harmonic potentials is increased 
$T_0(\mu)$ is decreased at the expense of lateral
transmissions $T_{-1,1}(\mu)$ and the three transmission
$T_{0,-1,1}(\mu)$ behaves in a complementary manner and are
bounded. While the sum of the three transmission $T_{0,-1,1}(\mu)$
converge towards unity, as required by the unitarity condition.

\section*{Acknowledgments}

The generous support provided by the Saudi Center for Theoretical Physics (SCTP)
is highly appreciated by all authors. HB and AJ acknowledges partial support
by King Fahd University of petroleum and minerals under the theoretical physics
research group project RG1306-1 and RG1306-2.

\end{document}